\documentclass[a4paper, 12pt]{article}

\usepackage{natbib}

\usepackage{graphicx}
\usepackage{amsfonts}
\usepackage{latexsym}
\usepackage{amsmath, amsthm, amssymb}
\usepackage{setspace}
\usepackage{url}
\usepackage{multirow}
\usepackage{multicol}
\usepackage{color}
\usepackage{psfrag}

\newtheorem{lem}{Lemma}

\usepackage[margin=3cm]{geometry}

\newcommand{\R}{\mathbb{R}}
\newcommand{\ty}{\tilde{y}}
\newcommand{\bz}{\mathbf{z}}
\newcommand{\sz}{z^*}
\newcommand{\bsz}{\bz^*}
\newcommand{\bY}{\mathbf{Y}}
\newcommand{\sY}{Y^*}
\newcommand{\bsY}{\mathbf{Y^*}}

\newcommand{\ep}{\epsilon}
\newcommand{\sig}{\sigma}
\newcommand{\hf}{\hat{f}}

\newcommand{\bu}{\mathbf{u}}
\newcommand{\bl}{\mathbf{l}}
\newcommand{\by}{\mathbf{y}}
\newcommand{\p}{\mathbb{P}}
\newcommand{\cN}{\mathcal{N}}

\title{Multiscale interpretation of taut string estimation and
its connection to Unbalanced Haar wavelets}
\author{
Haeran Cho and Piotr Fryzlewicz
\thanks{Department of Statistics, Columbia House, London School of Economics, Houghton Street, London, WC2A 2AE, UK.
\protect \ E-mail: {\tt h.cho1@lse.ac.uk, p.fryzlewicz@lse.ac.uk}
}
}

\begin{document}

\maketitle

\begin{center}
\textbf{Abstract}
\end{center}

We compare two state-of-the-art non-linear techniques for nonparametric function estimation via piecewise constant approximation: the taut string and the Unbalanced Haar methods. 
While it is well-known that the latter is multiscale, it is not obvious that the former can also be interpreted as multiscale. 
We provide a unified multiscale representation for both methods, which offers an insight into the relationship between them as well as suggesting lessons both methods can learn from each other.

\vspace{10pt} \textbf{keywords:} multiscale, Unbalanced Haar wavelets, taut string, nonparametric function estimation.

\section{Introduction}
\label{sec:intro}

A canonical problem in nonparametric regression is the estimation of a one-dimensional function $f$ from noisy observations $y$ in the additive model
\begin{eqnarray}
y_t=f\left(\frac{t}{n}\right)+\ep_t, \ t=1, \cdots, n,
\label{model}
\end{eqnarray}
where the observations $\{y_t\}_{t=1}^n$ are taken on an equispaced grid. 
In the simplest version of (\ref{model}), the noise $\{\ep_t\}_{t=1}^n$ is assumed to be iid Gaussian, which is not necessarily a realistic assumption in many applied problems, but serves as an excellent benchmark for comparing estimation techniques and gauging their potential performance in more complex models, in the sense that if a method misperforms in the model (\ref{model}) with iid Gaussian noise, there is normally little chance of it performing well in more complex settings.

In particular, the problem of estimating $f$ using piecewise constant estimators
has attracted considerable attention.
The class of piecewise constant functions is flexible in approximating a wide range of function spaces (e.g. see \cite{devore1998}).
Also, piecewise constant estimates are easy to interpret, as breakpoints in the
estimate indicate significant changes in the mean of the data, while the constant intervals between the breakpoints represent regions where the mean remains approximately the same.
It is well-known that if the underlying function $f$ is spatially inhomogeneous, non-linear piecewise constant estimators perform better than linear estimators. 
Therefore, in what follows, we discuss non-linear approaches.

Without attempting to be exhaustive, we mention a few recent, well-performing
estimation techniques. 
Wavelet thresholding estimation was first introduced in \cite{donoho1994}, where the thresholded estimator was shown to be theoretically tractable and perform well.
By using Haar wavelets, piecewise constant estimators are obtained.
The CART methodology \citep[Classification and Regression Trees]{breiman1983} performs greedy binary splitting to grow a partition, whose terminal nodes yield a piecewise constant estimator. 
In \citet{engel1997}, a method for locally adaptive histogram construction was introduced, which is based on a tree of dyadic partitions and hence obtains a multiscale, piecewise constant estimator.
\cite{polzehl2000} presented Adaptive Weight Smoothing, a data-driven local averaging procedure with an adaptive choice of weights, which iteratively 
produces a piecewise constant estimator.
More recently, methods involving the complexity-penalized likelihood optimization were proposed for estimating an unknown function by piecewise polynomials \citep{comte2004,kolaczyk2005}, which can be adopted to produce piecewise constant estimators.

In this paper, we are particularly interested in two methodologies, the Unbalanced Haar (UH) technique \citep{piotr2007} and the taut string (TS) based estimation (see e.g. \cite{barlow1972} and \cite{davies2001}). 
Both techniques are computationally fast, achieve theoretical consistency, and exhibit excellent performance in numerical simulation studies. 
The former involves the decomposition of the data with respect to orthonormal 
Haar-like basis vectors with jumps not necessarily in the middle of their support, while the latter finds a piecewise constant estimator via penalizing its total variation.

Our aim in this paper is to compare these two methods and discover links between them.
The UH technique is multiscale by nature \citep{piotr2007}, yet the multiscale character of the TS technique is less obvious, and has not been noted in the literature before.
Thus, in order to establish links between the two methods, we first provide an interesting multiscale interpretation of the TS technique. 
This then enables us to better understand similarities and differences between the UH and TS techniques, and establish a unified estimation methodology, which both the UH and the TS technique are instances of. 
Finally, taking advantage of this common framework, we derive lessons which either method can learn from the other.

The rest of paper is organized as follows. 
In Section \ref{sec:both}, we provide a description of the UH and TS techniques, as well as flowcharts of their algorithms, which offer an insight into the relationship between their physical interpretations.
Then follows the comparison study, including the understanding of the two techniques in the framework of breakpoint detection (Section \ref{sec:comp:study}). 
We conclude the discussion in Section \ref{sec:lesson} by listing some ways of 
possible improvement and extension for both techniques, which suggest avenues for further research.

\section{Unbalanced Haar and taut string techniques}
\label{sec:both}

In this section, we give an overview of the UH and TS techniques.
In particular, we emphasize the explicit multiscale nature of the UH methodology.
One contribution of this paper is to cast a new light on the TS technique via its new multiscale interpretation, which is achieved by introducing multiscale algorithms for both methods in Section \ref{sec:flowchart}.
These new algorithms are key to understanding and comparing the two techniques.

\subsection{Unbalanced Haar technique}
\label{sec:simple:uh}

The UH technique consists of three steps: the transformation of $\{y_t\}_{t=1}^n$ with respect to an adaptively chosen UH wavelet basis, hard-thresholding of the wavelet coefficients, and the inverse UH transformation of the thresholded coefficients to yield an estimate of $f$. 
For the principles of traditional wavelet thresholding estimation (without the adaptive basis selection), the reader is referred to \cite{vidakovic1999}.

The UH wavelet basis vectors were first studied in \cite{girardi1997} as an extended version of classical Haar wavelet vectors, the extension being that the breakpoint was permitted to occur anywhere within their support.
Let $s$ and $e$ denote the start and end of a generic interval, respectively, and let $b$ denote the location of the breakpoint.
Then, a UH vector on the interval $[s, e]$ with breakpoint $b$, $\psi_{s, b, e}$, is defined as $\psi_{s, b, e}(l)=$
$
\left\{\frac{1}{b-s+1}-\frac{1}{e-s+1}\right\}^{1/2}\mathbb{I}_{[s, b]}(l)
-\left\{\frac{1}{e-b}-\frac{1}{e-s+1}\right\}^{1/2}\mathbb{I}_{[b+1,e]}(l)$, for $\ s \le l \le e$.
Classical Haar wavelet vectors are a special case with $b=(s+e-1)/2$.

Denote the vector of observations as $\ty=(y_1, \ldots, y_n)^T$ and its sub-vector on a generic support $\{s,\ldots, e\}$ as $\ty_{s}^e=(y_s, \ldots, y_e)^T$. 
Noting that on a given support, the choice of breakpoints $b$ defines the choice 
of a UH basis, one way of UH basis selection is presented in \cite{piotr2007}. 
The first breakpoint $b_{1, 1}$ is chosen from $\{1, \ldots, n\}$ such that 
the inner product between $\ty$ and $\psi_{1, b_{1, 1}, n}$ is maximized in absolute value, i.e. $b_{1, 1}=\arg\max_{b\in\{1, \ldots, n\}}\left|\langle \ty, \psi_{1, b, n}\rangle\right|$. 
The explicit expression for the UH wavelet coefficient is given in (\ref{uh:loc}).
The next breakpoints are chosen similarly on the supports defined by the previously chosen breakpoint, $\{1, \ldots, b_{1, 1}\}$ and $\{b_{1, 1}+1, \ldots, n\}$, and the same procedure is repeated until it is no longer 
possible to divide any support into two.
Then $\ty$ is transformed with respect to the orthonormal basis defined
by the selected breakpoints. 
The next step is the hard-thresholding of the wavelet coefficients by 
setting to zero those which fall below the universal threshold 
$\sig\sqrt{2\log n}$. 
In practice the standard deviation of the noise is unknown but can 
be estimated as the median of the sequence 
$\{|y_{t+1}-y_{t}|/\sqrt{2}\}_{t=1}^{n-1}$ divided by the 
0.75-quantile of the standard normal distribution (which is approximately
equal to 0.6745). 
Finally the inverse transform is taken to obtain the final estimate 
$\hf^{UH}$, which is shown to be a mean-square consistent estimator 
for a wide range of functions, uniformly over those UH bases (however
they have been selected) which are not ``too unbalanced'' in the sense
that each basis vector should satisfy
\begin{eqnarray}
\max\left\{\frac{b-s+1}{e-s+1}, \frac{e-b}{e-s+1}\right\}\le p,
\label{constraint}
\end{eqnarray} 
for a fixed $p\in[1/2, 1)$. Thus, in practice, the maximisation of the
inner products as described above is performed in such a way that each time,
the maximum is only taken over those wavelets which satisfy condition
(\ref{constraint}), to ensure mean-square consistency of the resulting 
estimator.

We note that at the outset of the UH basis selection procedure, the entire observation vector is scanned in the search for $b_{1, 1}$, but then the scope of the search is iteratively narrowed down as each ``parent'' vector of observations gets iteratively divided into two ``children'', i.e. subvectors to the left and to the right of the previously detected breakpoint. 
Because 
of this natural  ``parent-child'' structure of the search, the UH estimation technique can be viewed as multiscale.

The recursive, binary nature of the UH technique shows its connection to the CART methodology.
However, the UH technique is more than a binary decision tree;
its key ingredient is that it furnishes a decomposition of data into 
wavelet coefficients, which can then be processed further depending on the 
aim of the analysis. In other words, the user of the UH methodology can enjoy 
the benefits of it being a wavelet technique, including generalizations to other (smoother) wavelets.

We also note that the binary decision tree is only one, ``top-down'', way of choosing a UH basis. Another way, which can be seen as a ``bottom-up'' approach, was 
introduced in \citet{piotr2007}. However, even the top-down UH estimator and CART 
differ significantly in that the former employs the usual universal wavelet 
thresholding, whereas the latter employs a ``hereditary'' mechanism whereby further 
subdivision is stopped as soon as a subinterval is judged to be a node.
An interesting connection between the dyadic (i.e. balanced) Haar approach 
and dyadic CART is given in \citet{donoho1997}, where again, it is noted that 
the dyadic CART estimator differs from the Haar thresholding estimator due to the 
heredity rule imposed on the tree structure.

In Section \ref{sec:flowchart}, we provide a more physical interpretation of the 
UH technique along with its flowchart representation.

\subsection{Taut strings}
\label{sec:simple:ts}

The TS technique is introduced in \cite{barlow1972} in the context of isotonic function estimation.
In the more general model (\ref{model}), it solves a penalized least squares functional where the 
penalty is based on the total variation norm \citep{mammen1997,davies2001}.
That is, it searches for a $\hf^{TS}$ satisfying 
\begin{eqnarray}
\hf^{TS}=\arg_{f}\min\left\{
\Vert \mathbf{f}-\by\Vert_2^2+\gamma\sum_t|f_{t+1}-f_t|\right\}, 
\label{ts:minimize}
\end{eqnarray}
where $\gamma$ is a tuning parameter. 
This is guaranteed to return a piecewise constant estimate whose number of breakpoints is a non-increasing function of $\gamma$.

One way of describing the computation of $\hf^{TS}$ is using the following ``string'' and ``tube'' arguments, which is referred to as the ``uniscale TS algorithm'' throughout this paper.
Denote the integrated process of observations $\{y_t\}_{t=1}^n$ as $\bY:=\{Y_t\}_{t=1}^n$, i.e. $Y_t=\sum_{u=1}^ty_u$ with $Y_0=0$.
Then imagine the graph of $\bY$ on the interval $[0, 1]$ which connects $\{(t/n, Y_t), 0 \le t \le n\}$, and also a tube of radius, say $\lambda>0$ (where $\lambda$ is related to the penalty constant $\gamma$ from (\ref{ts:minimize}) as $\gamma=2\lambda$), surrounding the graph $\bY$.
The tube consists of the lower bound $l_t:=Y_t-\lambda$ and the upper bound $u_t:=Y_t+\lambda$.
Then, suppose there is a string connecting $(0, Y_0)$ and $(1, Y_n)$, while being constrained to lie within the 
tube, and it is now pulled until it is taut, touching the tube on either side at possibly multiple ``knots''.
In other words, the taut string has the smallest length among functions 
$\{f:[0, 1]\rightarrow\R; f_0=Y_0, f_n=Y_n, l_t\le f_t\le u_t\}$, and its derivative 
coincides with $\hf^{TS}$ \citep{davies2001}.

Note that between two knots at which the string only touches the upper bound, it coincides with the greatest convex minorant (GCM) of $\bu$.
Similarly, between two knots where the string only touches the lower bound, it is the least concave majorant (LCM) of $\bl$.
Finally where the string switches from touching $\bu$ to touching $\bl$, a local maximum occurs in its derivative, and a local minimum occurs in the opposite manner.

Combined with a multiresolution bound over the empirical residuals, the TS technique is adopted in \cite{davies2001} for nonparametric regression with emphasis on consistent estimation of the number and locations of local extremes.
The authors propose a taut string algorithm which simultaneously computes the GCM of $\bu$ and the LCM of $\bl$ and finds the knots from left to right. 
In Section \ref{sec:flowchart}, we provide an alternative algorithm, accompanied by a flowchart, which reveals the multiscale nature of the taut string method. 
It is this multiscale interpretation of the taut string algorithm through which we derive the similarities and differences between the UH and TS techniques in Section \ref{sec:comp:study}.

\subsection{Unified multiscale description of UH and TS algorithms}
\label{sec:flowchart}

In introducing the flowcharts of the UH and TS techniques, we revisit the concept of a string and its knots.
Using the same notation as in Section \ref{sec:simple:ts}, consider a string, denoted by $\bz$, which connects $(0, Y_0)$ and $(1, Y_n)$ with a straight line.
We note that the algorithm for the UH technique is established in an \emph{adjusted} $y$-axis: 
we define a multiplying factor $\rho^{UH}$ on $t\in[s, e)$ as
\begin{eqnarray}
\rho^{UH}(t; s, e)=\sqrt{\frac{e-s+1}{(t-s+1)(e-t)}},
\label{uh:factor}
\end{eqnarray}
which adjusts the string and the integrated process to yield
$\sz_t=\rho^{UH}(t; s, e)\cdot z_t$ and $\sY_t=\rho^{UH}(t; s, e)\cdot Y_t$.
The adjusting factor $\rho^{UH}$ comes from the UH wavelet basis used to compute the wavelet coefficient.
It is designed such that the wavelet coefficient defined on the segment $[s, e]$ with a 
breakpoint at $t$ is equal to the product of $\rho^{UH}$ and the differential term between 
the local sum ($\sum_{u=s}^ty_u=Y_t-Y_{s-1}$) and the scaled global sum 
($\frac{t-s+1}{e-s+1}\cdot(Y_e-Y_{s-1})$) of the observations, see (\ref{uh:loc}) for further details.

\begin{figure}[htbp]
\centering
\includegraphics[width=1\textwidth, height=.75\textheight]{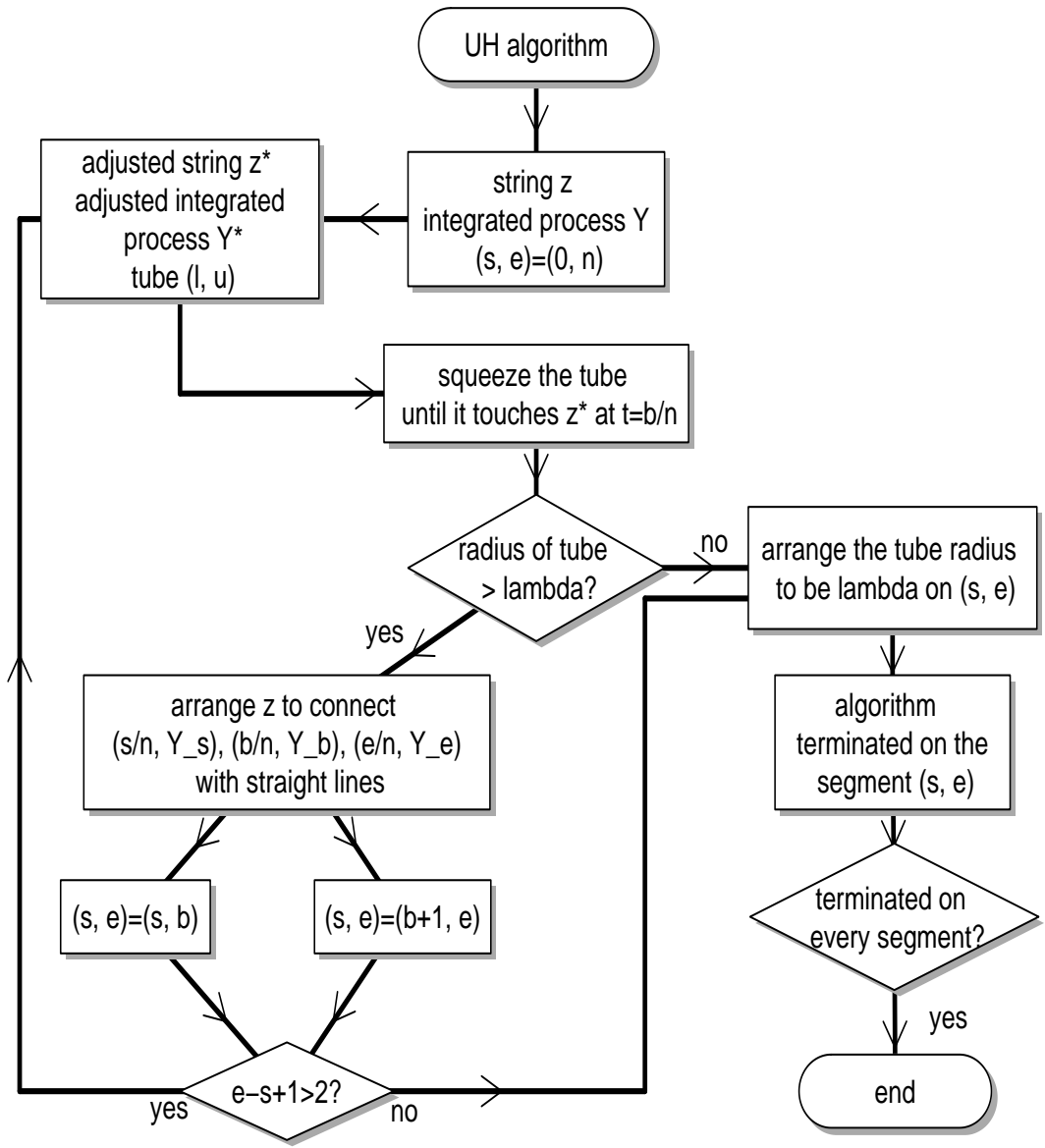}
\caption{\footnotesize{Flowcharts of UH algorithm.}}
\label{fig:uh:alg}
\end{figure}

\begin{figure}[htbp]
\centering
\includegraphics[width=1\textwidth, height=.75\textheight]{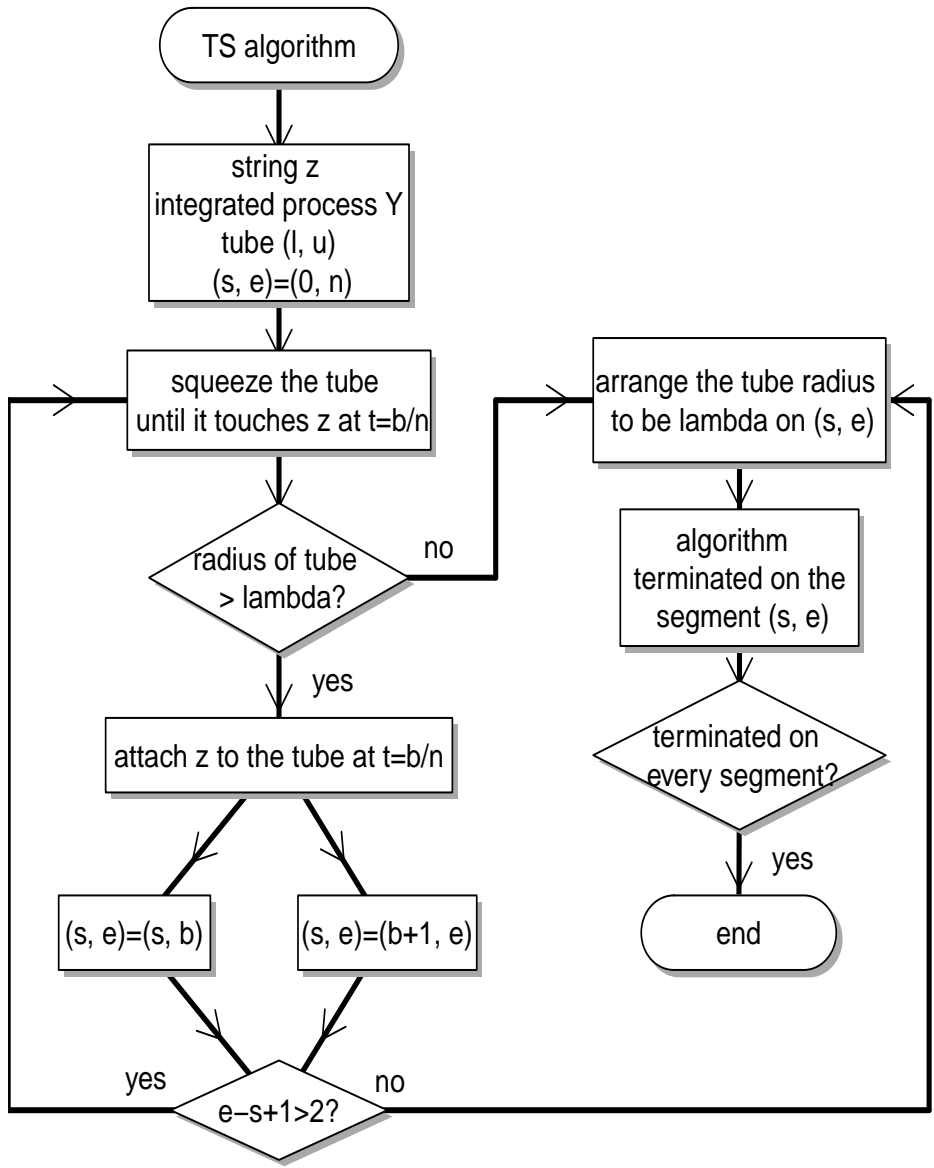}
\caption{\footnotesize{Flowcharts of TS algorithm; see Section \ref{sec:comp:study} for the comparison study between the UH and TS techniques.}}
\label{fig:ts:alg}
\end{figure}

Next, consider a tube of radius $r$ surrounding the integrated process $\bY$ (or its adjusted version $\bsY$ in the UH technique); however this time the radius is chosen to be so large that the string $\bz$ ($\bsz$) does not touch the tube. 
With this starting set-up, our algorithmic interpretation of the two techniques is summarized in the flowcharts in Figures \ref{fig:uh:alg}--\ref{fig:ts:alg}.


The two algorithms proceed similarly by squeezing the tube and re-arranging the string simultaneously.
By squeezing the tube, the first knot is detected at, say $t=b/n$, as the point where the tube first touches $\bz$ ($\bsz$).
If the radius of the squeezed tube is greater than a pre-specified value $\lambda>0$, the string is re-arranged (see point (ii) below) and two segments are defined by the knot at $t=b/n$, i.e. $(0, b/n)$ and $(b/n, 1)$. 
The same knot detection and string re-arrangement steps are repeated on each segment separately, as long as (a) the length of the segment is large enough for further division of the segment to be possible in the next iteration, and (b) the squeezed tube radius is greater than $\lambda$ on the given segment. If, on any segment, the radius of the tube is found to be less than $\lambda$, we set it to $\lambda$.
The estimation procedure is finished once the progression of the algorithm is terminated on every segment, and the estimator is obtained as the derivative of the string $\bz$.
In both algorithms, the current ``parent'' segment is always split into two ``children'' subsegments.
Therefore the same procedure is applied to the data at multiple scales, and thus we can conclude that not only the UH technique but also the TS technique is multiscale. 

While the basic steps of the two algorithms are similar (as described above), they differ in the following details.
\begin{itemize}
\item[(i)] The UH algorithm is performed in the adjusted $y$-axis, while the TS algorithm is performed in the original $y$-axis.
\item[(ii)] When a knot is detected with the squeezed tube having its radius larger than $\lambda$, the string is re-arranged differently; 
on a generic segment $(s/n, e/n)$, the UH algorithm arranges $\bz$ to connect $(s/n, Y_s)$ and $(b/n, Y_b)$ with a straight line, as well as $(b/n, Y_b)$ and $(e/n, Y_e)$ with a straight line; 
on the other hand, the TS algorithm attaches $\bz$ to the tube at the detected knot 
and further squeezing of the tube is applied with $\bz$ still being attached to it.
Note that the tube remains a symmetric band around the integrated process $\bY$ throughout the algorithm.
However, since $\bz$ consists of straight lines connecting two neighbouring knots 
(including $(0, Y_0)$ and $(1, Y_n)$), the slope of each line changes constantly as 
the radius of the tube decreases, and as a result, it is a constantly changing function on $[0, 1]$.
The attachment of $\bz$ to the tube can be observed in Figure \ref{fig:toy:ts}, where 
the upper right and lower middle figures show the state in between the detection of two knots.
In summary, our TS algorithm returns its estimator as the derivative of the taut string which is attached to the tube of radius $\lambda$ at zero, one or multiple knots and connects neighbouring knots with straight lines.
\end{itemize}

As opposed to the uniscale TS algorithm presented in Section \ref{sec:simple:ts}, the TS algorithm from our unified approach is referred to as the ``multiscale TS algorithm'' throughout the paper.
We emphasize that the multiscale TS algorithm returns exactly 
the same estimator as that obtained from the uniscale TS algorithm, and thus it also solves the penalized least squares problem in (\ref{ts:minimize}).
While applying the multiscale TS algorithm, when the first knot is detected with the tube squeezed so that its radius equals $\lambda_1$, the string in that state is equal to the string from the uniscale TS algorithm with the tube radius equal to $\lambda_1$. Then recursively applying the same argument, it can be seen that the multiscale TS algorithm produces exactly the same state of the tube and the string as the uniscale TS algorithm.

We note that the UH algorithm as presented in the flowchart (Figure \ref{fig:uh:alg}) is a modification of the description in Section \ref{sec:simple:uh}. 
The modification simplifies the graphical representation as well as the comparison between two techniques. 
In the flowchart, the algorithm terminates on a segment if the squeezed tube radius is smaller than $\lambda$, while the original algorithm terminates only when the length of the segment is too small (but then applies thresholding with the threshold set equal to $\lambda$).
This difference can affect the adaptivity of the final estimate $\hf^{UH}$ depending on the shape of underlying function $f$, and is further discussed in Section \ref{sec:lesson}.
We also note that the algorithm in Figure \ref{fig:uh:alg} does not take into account the condition imposed in (\ref{constraint}) when selecting $b\in(s, e)$, 
unlike the original UH algorithm as proposed in \citet{piotr2007}. 
However, this condition can easily be incorporated in both UH and TS algorithms and is only omitted for the simplicity 
of presentation.

\begin{figure}
\begin{minipage}[b]{1\linewidth}
\centering
\includegraphics[width=.7\textwidth, height=.45\textheight]{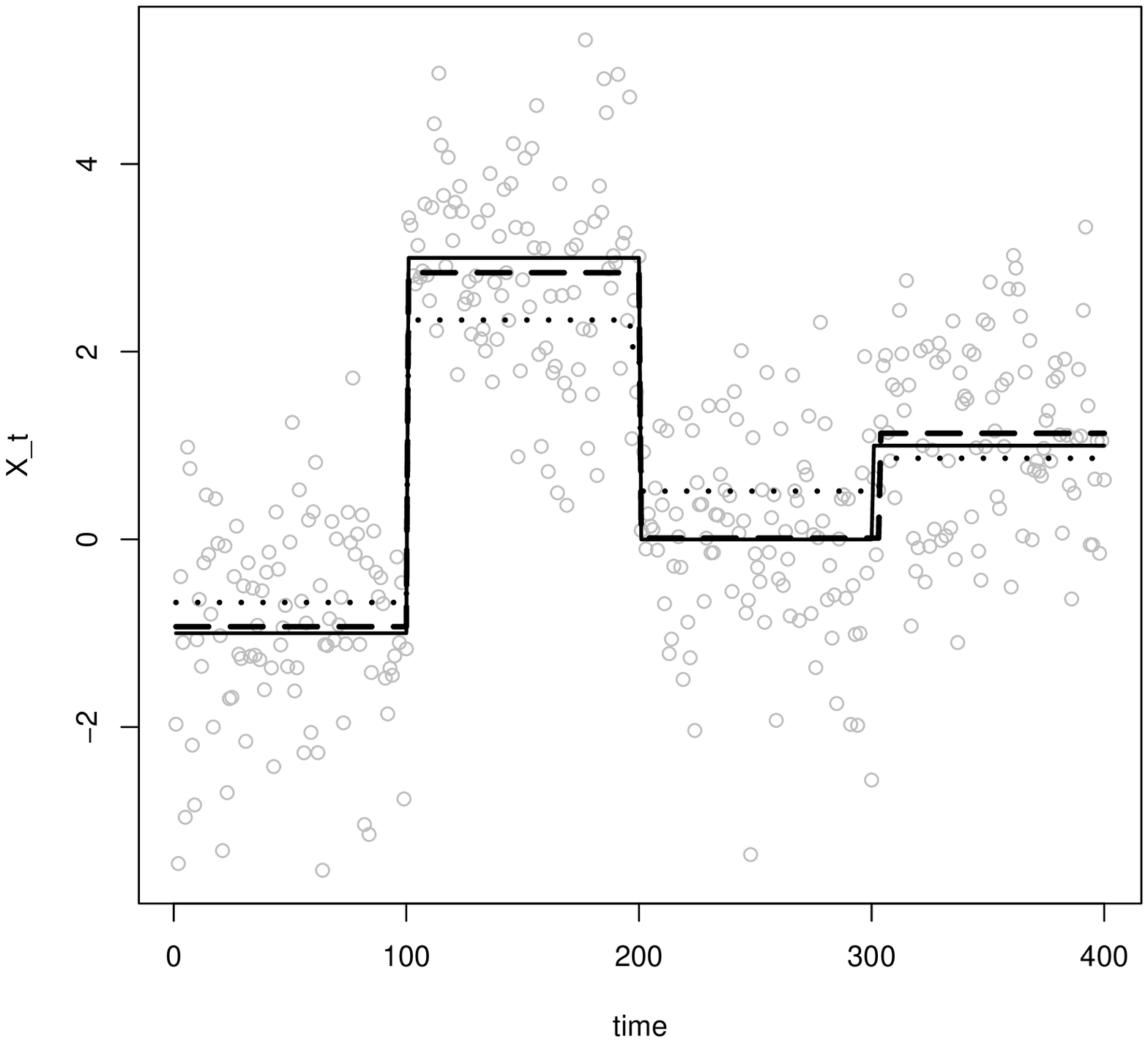}
\caption{\footnotesize{A toy example; $y_t$ (dots), $f$ (solid),  $\hf^{UH}$ (dashed), $\hf^{TS}$ (broken).}} 
\label{fig:toy}
\end{minipage}
\begin{minipage}[b]{1\linewidth}
\centering
\includegraphics[width=1\textwidth, height=.35\textheight]{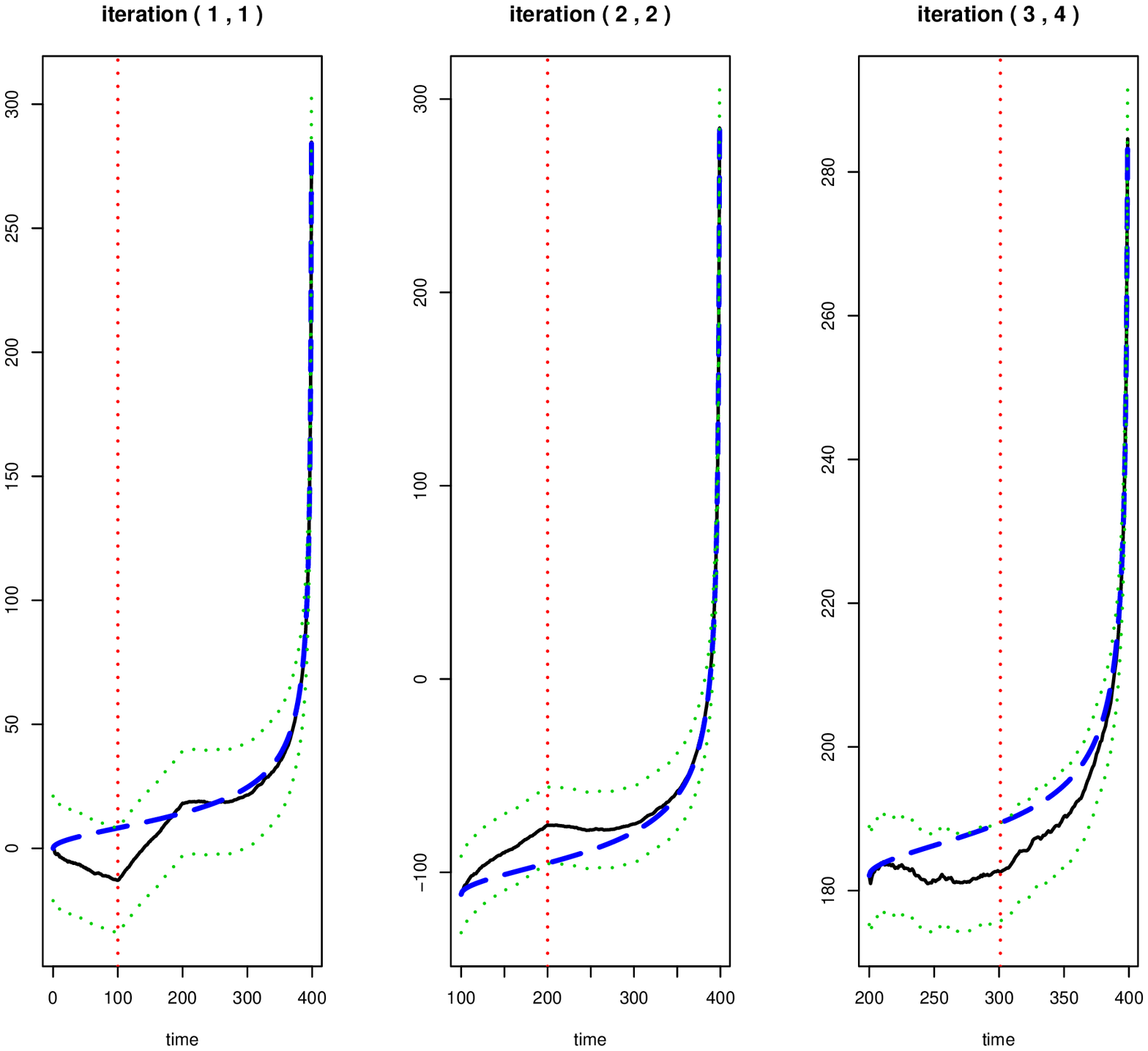}
\caption{\footnotesize{An application of UH algorithm to the model in Figure
\ref{fig:toy}; adjusted integrated process (solid), string (broken), tube (dotted), and the locations of the knots (vertical, dotted)}}
\label{fig:toy:uh}
\end{minipage}
\end{figure}

\begin{figure}
\begin{minipage}[b]{1\linewidth}
\centering
\includegraphics[width=1\textwidth, height=.7\textheight]{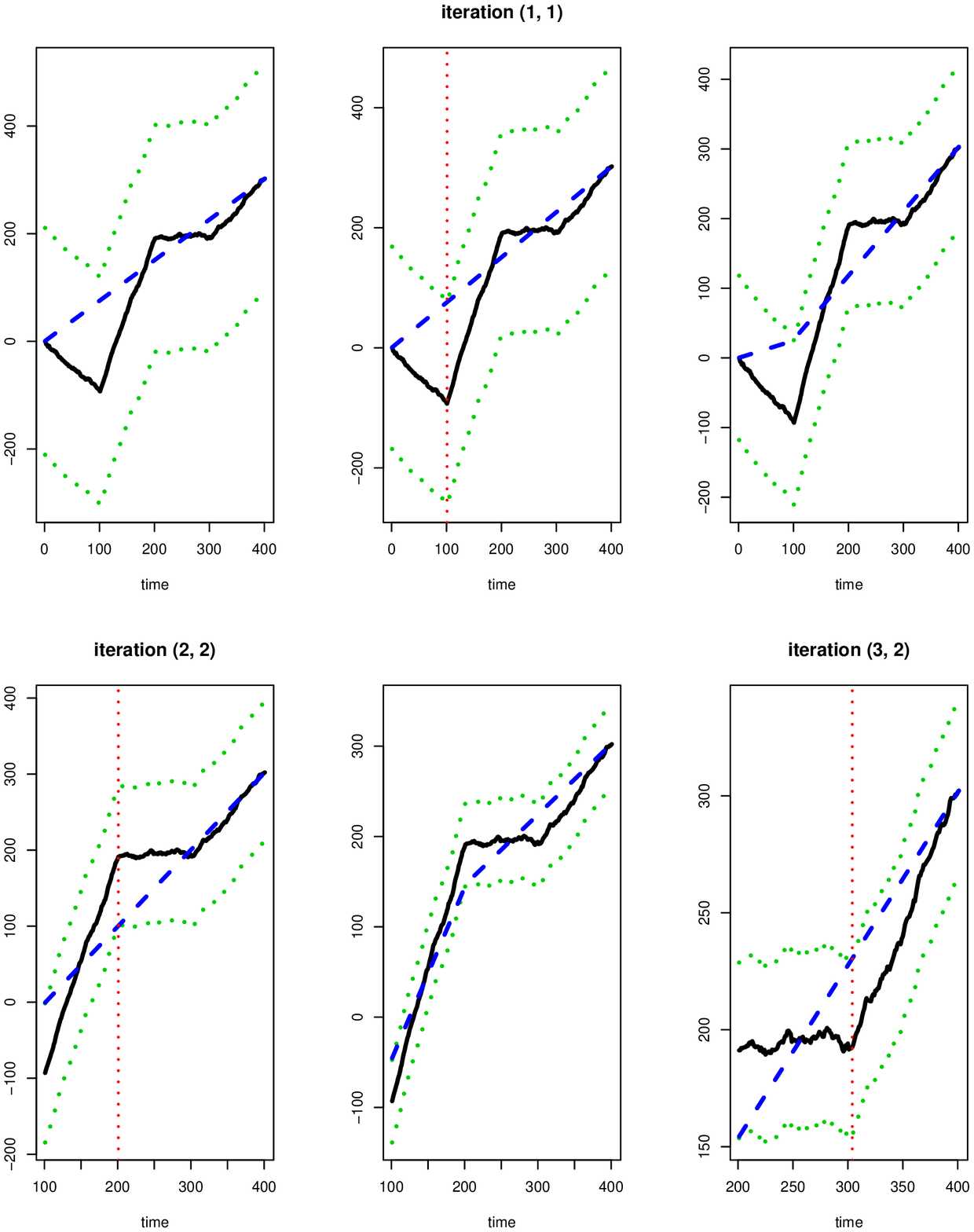}
\caption{\footnotesize{An application of TS algorithm to the model in Figure
\ref{fig:toy}; integrated process (solid), string (broken), tube (dotted), and the locations of the knots (vertical, dotted)}; the upper left figure shows the state of the tube and string at the beginning of algorithm; the upper right and lower middle figures show the state in between the detection of knots.}
\label{fig:toy:ts}
\end{minipage}
\end{figure}

We conclude this section by showing, in Figures \ref{fig:toy:uh}--\ref{fig:toy:ts}, iteration-by-iteration progression of both algorithms from our unified approach as applied to the toy example from Figure \ref{fig:toy}.
Iteration $(j, k)$ indicates that the knot is detected in the $j$th iteration on the $k$th segment from the left.

\section{Comparison of UH and TS techniques}
\label{sec:comp:study}

Based on the multiscale algorithms established above, we now provide a detailed comparison study between the two techniques.
Firstly, in Section \ref{sec:analytic}, we define the ``locating'' functions for both techniques, which are used to find the locations of knots in a given segment. 
The comparison study continues in Section \ref{sec:comp:theory} in the framework of breakpoint detection, which provides an insight into reasons why the UH and TS techniques often perform differently.

\subsection{Locating functions of UH and TS techniques}
\label{sec:analytic}

In the UH technique, the selection of a UH basis on a generic interval $[s, e]$ involves the computation of the inner product between $\ty_s^e$ and a set of UH wavelet vectors $\psi_{s, t, e}$ for $t\in(s, e)$.
The break in a wavelet vector, $b$, corresponds to the knot on the segment $(s/n, e/n)$ in the UH algorithm, and it is located as
\begin{eqnarray}
b&=&\arg\max_{t\in(s, e)}\left\vert\langle \ty_s^e, \psi_{s, t, e} \rangle\right\vert \nonumber \\
&=&\arg\max_{t\in(s, e)}\left\vert \sqrt{\frac{e-t}{(e-s+1)(t-s+1)}}\left(Y_t-Y_{s-1}\right)
-\sqrt{\frac{t-s+1}{(e-s+1)(e-t)}}\left(Y_e-Y_t\right) \right\vert \nonumber \\
 &=&\arg\max_{t\in(s, e)}\left\vert
\sqrt{\frac{e-s+1}{(t-s+1)(e-t)}}\left\{
\frac{t-s+1}{e-s+1}\left(Y_e-Y_{s-1}\right)-\left(Y_t-Y_{s-1}\right)
\right\}\right\vert \nonumber \\
&=&\arg\max_{t\in(s, e)} c^{UH}(t; s, e).
\label{uh:loc}
\end{eqnarray}

$c^{UH}(b; s, e)$ can be seen as the radius of the tube in its adjusted $y$-axis when it touches the string at $b/n$, as well as having the interpretation of being the UH wavelet coefficient of $\ty_s^e$ in absolute value.
Therefore the step comparing the squeezed tube radius to $\lambda$ is equivalent to the hard-thresholding of the wavelet coefficients and it justifies setting $\lambda$ equal to the universal threshold.

We now derive the locating function for the TS algorithm.
Conditional on the string touching the tube at time $t$, let $g_t$ indicate whether it touches its upper ($g_t=1$) or lower ($g_t=-1$) bound.
Initially, as the bounds of the tube approach the string, we note that the first knot is chosen as
\begin{eqnarray}
b=\arg_{t\in(0, n)}\max_{g_t=\pm
1}g_t\cdot\left(\frac{t}{n}Y_n-Y_t\right). \label{ts:loc:first}
\end{eqnarray}
With the convention that $g_0=g_n=0$, further knots on a generic interval $(s, e)$ are located as
$b=\arg_{t\in(s, e)}\max_{g_t=\pm 1} c^{TS}(t; s, e)$, where
\begin{eqnarray*}
c^{TS}(t; s, e)=\left\{\begin{array}{l}
g_t\cdot\left\{\frac{t-s+1}{e-s+1}\left(Y_e-Y_{s-1}\right)-\left(Y_t-Y_{s-1}\right)\right\}
\mbox{ if } g_{s-1}=g_e,
\\
\frac{e-s+1}{(e-s+1)(g_t-g_{s-1})-(t-s+1)(g_e-g_{s-1})}
\left\{\frac{t-s+1}{e-s+1}\left(Y_e-Y_{s-1}\right)-\left(Y_t-Y_{s-1}\right)\right\}
\\
\mbox{if } g_{s-1}\ne g_e.
\end{array}\right.
\end{eqnarray*}

Comparing $c^{UH}$ and $c^{TS}$ shows that the two methods can be regarded as both ``integrated'' and ``differential'' in the sense
that they are applied to the integrated process $\bY$ (up to the adjusting factor $\rho^{UH}$ for the UH 
estimator) and their test statistics share the term $\left(\frac{t-s+1}{e-s+1}\left(Y_e-Y_{s-1}\right)-(Y_t-Y_{s-1})\right)$, 
the differential term between the scaled global sum and the local sum on a given segment.
To see the difference between their multiplying factors, we quote the following lemma from  \cite{venkatraman1993}.
In our context, Lemma \ref{lem:one} implies that when signal $f$ is piecewise constant and there is 
no noise in the observations, the maximum of $c^{UH}$ is attained only at the true breakpoints of $f$ at every iteration of the UH algorithm.

\begin{lem}[Lemma 2.2 in \cite{venkatraman1993}]
\label{lem:one} Let $l>0$ be an integer and $0=a_0<a_1<\ldots<a_l<a_{l+1}=1$.
Assume a piecewise constant function $f(x)$ with breakpoints $a_i, \ i=1, \ldots, l$ and let $f(x)=\lambda_i$ for $x\in(a_{i}, a_{i+1}]$, where
$\lambda_i\ne\lambda_{i+1}$.
Finally we assume that $\sum_{i=0}^l(a_{i+1}-a_i)\lambda_i=0$.
Define the function $h^{UH}$ as
\begin{eqnarray}
h^{UH}(x)=\frac{\sum_{j=1}^i(a_j-a_{j-1})\lambda_{j-1}+(x-a_i)\lambda_i}{\sqrt{x(1-x)}},
\label{uh:f}
\end{eqnarray}
for $x\in[a_i, a_{i+1}]; 0 \le i \le l$.
Denote $h^*=\max_{x\in(0, 1)}\left|h^{UH}(x)\right|$ and $x^*$ as where the maximum value is attained, i.e. $h^{UH}(x^*)=h^*$.
Then there exists $1\le i \le l$ such that $a_i=x^*$, i.e., the maximum of $|h^{UH}|$ can only be attained at one of $a_i$'s.
\end{lem}

\begin{figure}[ht]
\centering
\includegraphics[width=1\textwidth, height=.45\textheight]{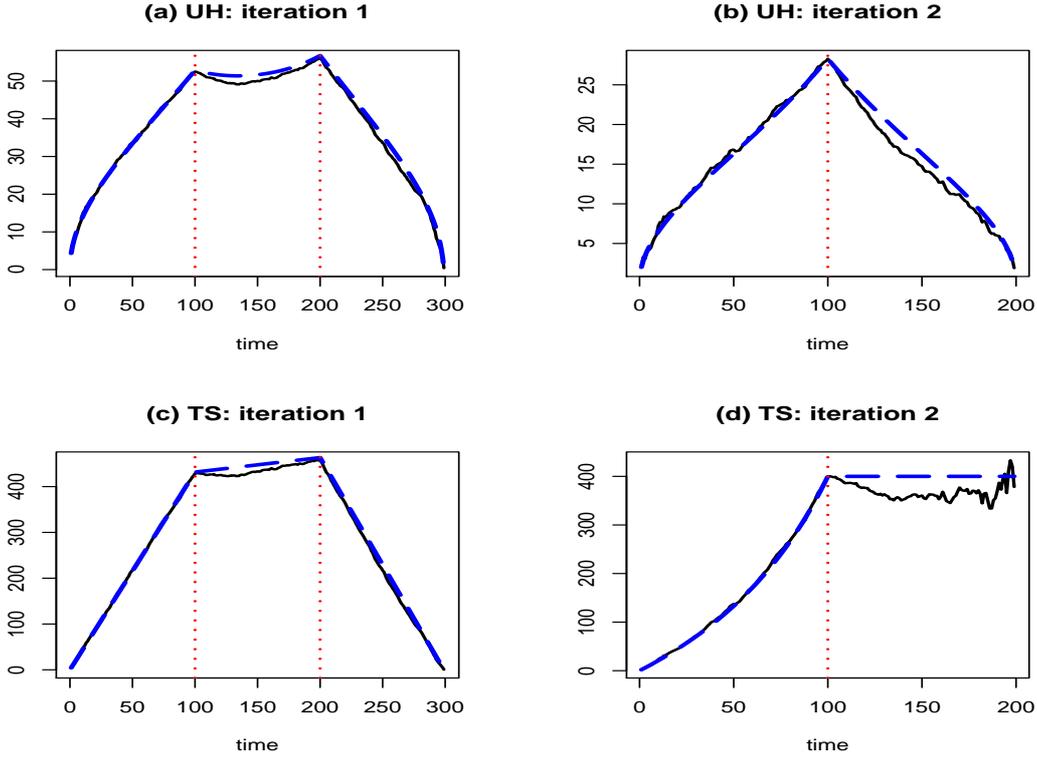}
\caption{\footnotesize{(a), (b) $\left|c^{UH}(t; s, e)\right|$ at iteration 1, 2; (c), (d) $\left|c^{TS}(t; s, e)\right|$ at iteration 1, 2; vertical dotted: true breakpoints, dashed: noiseless observations, solid: noisy observations.}} 
\label{fig:one}
\end{figure}

Simple algebra shows that $h^{UH}$ is equivalent to $c^{UH}$ for $x=t/n\in(0, 1)$. 
The equivalent of $h^{UH}$ for the TS technique is defined in the notation of Lemma \ref{lem:one} as
\begin{eqnarray}
&&h^{TS}(x)=\frac{\sum_{j=1}^i(a_j-a_{j-1})\lambda_{j-1}+(x-a_i)\lambda_i}
{\alpha_1x+\alpha_2(1-x)}
\label{ts:f} \\
&& \mbox{where \ } \alpha_k\in\{0, \pm 1, \pm 2\}; k=1, 2,
\mbox{ subject to } |\alpha_1+\alpha_2|=2,
\nonumber
\end{eqnarray}
for $x\in[a_i, a_{i+1}]; 0 \le i \le l$.
The particular values taken by $\alpha_1$, $\alpha_2$ depend on whether the string touches the lower or upper bound at the start and end of the segment defined by $[a_i, a_{i+1}]$.
Figure \ref{fig:one} shows interesting characteristics of the two locating functions, where the UH and TS algorithms are applied to both noiseless and noisy observations of (\ref{ex:one}) with $n=300$,
\begin{eqnarray}
f(u)=\left\{\begin{array}{ll}
-4 & \mbox{ for } u\in(0, 1/3], \\
0 & \mbox{ for } u\in(1/3, 2/3], \\
5 & \mbox{ for } u\in(2/3, 1].  \\
\end{array}\right. \label{ex:one}
\end{eqnarray}

First, consider the example with noiseless observations (dashed lines).
The upper panel shows $c^{UH}$ at first two iterations ($(s, e):(1, 300) \rightarrow (1, 200)$), where it is clear that the (local) maxima are attained exactly at the true breakpoints ($t=100$, $200$).
The lower panel shows $c^{TS}$ at first two iterations, where two different shapes of the locating function are observed. $c^{TS}$ is piecewise linear at the first iteration, and, at the second iteration, it reaches a plateau at $t=100$ and remains constant on $[100, 200)$. 
For a piecewise constant signal 
function $f$, either shape can occur depending on which side of the tube the string has been attached to in previous iterations, i.e. on the values of $g_s$, $g_e$, and $g_b$.

In either case, it is clear that $c^{TS}$ does not ``point out'' the locations of true breakpoints as distinctively as $c^{UH}$ does since the change in the derivative of $c^{TS}$ is not as dramatic as in that of $c^{UH}$ around each breakpoint.
Thus we conclude that there is no theoretical equivalent of Lemma \ref{lem:one} for $h^{TS}$.
This difference may lead to the TS estimate reflecting the true breakpoint
structure less accurately than the UH estimate.
 
\subsection{Link to breakpoint detection}
\label{sec:comp:theory}

A theoretical study of a family of test statistics for breakpoint detection is performed in \cite{brodsky1993}. 
Their study, in light of the relationship of these test statistics to $c^{UH}$ and $c^{TS}$, adds further strength to our arguments from the previous section.
In Chapter 3.5 of the book, the problem of \emph{a posteriori} (retrospective) 
breakpoint detection is considered, where the task is to find an abrupt change in the mean value of a random sequence. 
Let $\{x_t\}_{t=1}^n$ be a realization of a Gaussian process with at most one breakpoint in its mean and otherwise iid, and $X_t$ be the integrated process of $x_t$, i.e. $X_t=\sum_{u=1}^tx_u$. 
Then a family of test statistics indexed by $\delta$ was proposed as 
\begin{eqnarray}
d_{\delta}(t)=\left\{\frac{t}{n}\left(1-\frac{t}{n}\right)\right\}^{\delta}
\left\{\frac{1}{t}X_t-\frac{1}{n-t}\left(X_n-X_t\right)\right\},
\label{delta}
\end{eqnarray}
where $t\in\{1, \ldots, n\}$ and $\delta\in[0, 1]$. 
A breakpoint candidate is chosen as $\hat{b}_{\delta}=\arg\max_t|d_{\delta}(t)|$, and if $|d_{\delta}(\hat{b}_{\delta})|$ exceeds a test criterion, $\hat{b}_{\delta}$ becomes the estimated breakpoint. 
It can be shown with simple algebra that $d_{1/2}$ corresponds to $c^{UH}$, and $d_{1}$ to $c^{TS}$ (at the first iteration of the TS algorithm and each time when $g_s=g_e$ later on, i.e. when the string is attached to the same side of tube at $t=s$ and $t=e$).

Below we summarize the asymptotic results from \citet{brodsky1993} on the probabilities of type $1$ error (false alarm, i.e. the test statistic exceeding the test criterion although there is no breakpoint), type $2$ error (false tranquillity, i.e. the test statistic being smaller than the test criterion although there is a breakpoint), and the estimation error in the distance between the estimated and true breakpoints.
Note that the single breakpoint in the following (ii), (iii) is constrained to exist within $[a_1, a_2]$ 
where $0<a_1<a_2<1$, which is in accordance with the assumption (\ref{constraint}) for the UH technique.

\begin{itemize}
\item[(i)]
When there is no breakpoint present in the observations, the asymptotic rate of convergence for the probability of 
a type $1$ error increases in $\delta$, i.e. $d_1$ is asymptotically the best in not causing any false alarm.
\item[(ii)] When there is a single breakpoint, the asymptotic rate of convergence for the probability of a type 
$2$ error decreases in $\delta$, i.e. $d_0$ is asymptotically the best at detecting that there is a breakpoint.
\item[(iii)] When there is a single breakpoint, say $b$, the asymptotic rate of convergence for the estimation 
error probability $\p\left(\left\vert\hat{b}_{\delta}-b\right\vert>\xi\right)\longrightarrow 0$ is maximized 
when $\delta=1/2$, i.e., $d_{1/2}$ is asymptotically the best at estimating the location of the breakpoint.
\end{itemize}

Note that the above (i) and (ii) are obtained under the assumption that the same critical value is used for 
all $d_{\delta}(t), \delta\in[0,1]$. Then, for a fixed critical value, the rate of convergence for the probabilities 
of type $1$ and type $2$ errors are optimized when $\delta=0$ and $\delta=1$, respectively.

Suppose now that we choose the critical value $c_\delta$ (depending on $\delta$) such that the probability of 
a type $1$ error is fixed at $\alpha$.
Since the iid noise satisfies $\ep_t\sim\cN(0, 1)$, Theorem 3.5.1 of \citet{brodsky1993} implies that
\begin{eqnarray}
c_0=\sqrt{\frac{2A}{\Delta n}}, \ c_{1/2}=\sqrt{\frac{2A}{n}}, \mbox{ and } c_1=\sqrt{\frac{A}{2n}},
\label{critical}
\end{eqnarray}
where $A=-\log(\alpha)$ and $\Delta=\min(a_1(1-a_1), a_2(1-a_2))$.

With the above critical values, we can compare the rate of convergence at which the probability of 
a type $2$ error tends to $0$ for different choices of $\delta$.
Let $\beta_{\delta}(n)$ denote the probability of a type $2$ error for each $\delta$, $h$ be the 
magnitude of the jump at the breakpoint, and $p:=b(1-b)\le 1/4$. 
It is noted in \citet{brodsky1993} that when the critical value does not satisfy $c_\delta<hp^\delta$, 
the probability of a type $2$ error is positive for all $n$ and tends to $1$ as $n\rightarrow\infty$.
Therefore assuming $c_\delta<hp^\delta$, we obtain the following from their Theorem 3.5.2,
\begin{eqnarray}
\beta_{\delta}(n)\sim\exp\left(-\frac{n(hp^{\delta}-c_{\delta})^2}{2p^{2\delta-1}}\right)=
\exp\left(-\frac{nC_{\delta}}{2}\right).
\label{beta:rate}
\end{eqnarray}
By plugging in $c_{\delta}$ from (\ref{critical}), each $C_{\delta}$ is obtained as
\begin{eqnarray}
C_0=\left(h\sqrt{p}-\sqrt{\frac{2pA}{\Delta n}}\right)^2, \
C_{1/2}=\left( h\sqrt{p}-\sqrt{\frac{2A}{n}} \right)^2, \
C_1=\left(h\sqrt{p}-\sqrt{\frac{A}{2pn}}\right)^2.
\nonumber
\end{eqnarray}
Recalling that the true breakpoint (if it exists) satisfies $b\in[a_1, a_2]$, $p\ge\Delta$ and thus we have $2p/\Delta\ge 2$ and $1/(2p)\ge 2$.
Therefore 
$C_{1/2}\ge C_\delta$, $\delta=0, 1$, i.e. when the type $1$ error probability is fixed, the 
rate of convergence for probability of a type $2$ error is better for $\delta=1/2$ than for $\delta=0, 1$.

In the above sense, $c^{UH}$ is more alert at breakpoint detection, in detecting both its presence and 
its location, in comparison to $c^{TS}$. Combined with the observation made in Section \ref{sec:analytic}, 
when estimating a piecewise constant signal with the emphasis on breakpoint detection, it is 
likely that the UH technique would perform better than the TS technique.

\section{Possible lessons and directions for future research}
\label{sec:lesson}

While the comparison study between the UH and TS techniques is interesting in itself, it also provides, by establishing links between them, common ``ground'' on which the two methods can learn lessons from each other, potentially leading to new developments in the area of nonparametric function estimation.

\begin{description}
\item[Choice of threshold.] 
The UH algorithm uses the universal threshold $\sigma\sqrt{2\log n}$ as the critical radius $\lambda$.
By comparing the multiplying factors of $c^{UH}$ and $c^{TS}$, we can derive the corresponding critical radius for the multiscale TS algorithm. 
The equivalent of $\rho^{UH}$ for the multiscale TS algorithm, say $\rho^{TS}$, satisfies $\rho^{TS}(b; s, e)/\rho^{UH}(b; s, e)=C_{\alpha}\sqrt{e-s+1}$, 
where $\alpha=(b-s+1)/(e-s+1)$ and $C_{\alpha}$ is a constant depending on $\alpha$, $g_s$, $g_e$, and $g_b$. 
Therefore $C_\alpha\sigma\sqrt{2n\log n}$ can be used as the stopping radius in the multiscale TS algorithm.

\item[UH basis selection.] 
The mean-square consistency result given in \citet{piotr2007} holds for any UH basis as long as the breakpoint in each wavelet vector is not too ``unbalanced''. 
The TS algorithm provides yet another way of constructing a UH basis.

\item[Local squeezing.] 
To improve the convergence rate at local extremes, \cite{davies2001} combine the taut string technique with a multiresolution bound over estimated residuals, applying an additional local squeezing step to the taut string estimate.
It may be possible to derive a similar theoretical result on the estimated UH residuals $\by-\hf^{UH}$ and apply a similar local squeezing to obtain a sharper estimate.

Although it does not contain explicit local squeezing, the original UH algorithm as presented in \citet{piotr2007} obtains 
the UH wavelet decomposition down to the finest scale and then applies the thresholding of wavelet coefficients. 
This can be seen as a replacement for / equivalent of the local squeezing used in the TS 
technique, as it enhances the adaptivity of the UH estimator. Similar modification can readily be made to 
our version of the TS algorithm.

\item[Controlling the total variation.] 
The total variation penalty in (\ref{ts:minimize}) restricts the string to be attached to one of the bounds of the tube. 
Therefore by modifying the re-arrangement of the string in the UH algorithm, similar control over the total variation of the estimated function could be achieved.

\item[Extensions to non-Gaussian error distributions.]
In practice, the assumption of Gaussianity is violated in many nonparametric estimation problems, such as Poisson intensity or volatility estimation.
In \cite{dumbgen2009}, the extensions of taut strings are discussed under the assumption that the noise follows a distribution from the exponential
family.
Their final estimate is obtained by transforming $\hf^{TS}$, the estimate from the least squares setting in (\ref{ts:minimize}), via a known function.
The same arguments may be applied to $\hf^{UH}$ when the prior knowledge on the noise distribution is available.

On the other hand, for the cases where the exact form of the relationship between the mean and variance functions is unknown, a data-driven 
wavelet-based estimation technique is proposed in \cite{piotr2008}, where the use of UH wavelets is readily applicable. 
By treating the variance stabilization step as the adjustment of the $y$-axis, its extension to the TS technique is also feasible via applying an appropriate multiplying factor to the string and the integrated process.
\end{description}

\subsection*{Acknowledgements}

The authors are grateful to the Editor, Associate Editor and two Referees for their stimulating reports, which led to a significant improvement of this paper.

\bibliographystyle{spbasic}
\bibliography{mbib}

\end{document}